\documentclass[twocolumn,english,aps,prl,floats,showpacs]{revtex4}
\usepackage[T1]{fontenc}
\usepackage[latin9]{inputenc}
\usepackage{amsmath}
\usepackage{amssymb}
\usepackage{graphicx}
\usepackage{esint}

\makeatletter
\@ifundefined{textcolor}{}
{%
 \definecolor{BLACK}{gray}{0}
 \definecolor{WHITE}{gray}{1}
 \definecolor{RED}{rgb}{1,0,0}
 \definecolor{GREEN}{rgb}{0,1,0}
 \definecolor{BLUE}{rgb}{0,0,1}
 \definecolor{CYAN}{cmyk}{1,0,0,0}
 \definecolor{MAGENTA}{cmyk}{0,1,0,0}
 \definecolor{YELLOW}{cmyk}{0,0,1,0}
}

\makeatother

\usepackage{babel}
\begin{document}

\title{Collapse of Skyrmions in 2$d$ Ferro- and Antiferromagnets}

\author{Liufei Cai, Eugene M. Chudnovsky, and D. A. Garanin}

\affiliation{Physics Department, Lehman College, City University of New York \\
 250 Bedford Park Boulevard West, Bronx, New York 10468-1589, USA}

\date{\today}
\begin{abstract}
Collapse of a skyrmion due to the discreteness of a crystal
lattice in isotropic two-dimensional ferro- and antiferromagnets
has been studied analytically and by numerical solution of equations of motion for up to
2000$\times$2000 classical spins on a square lattice coupled via Heisenberg
exchange interaction. Excellent agreement between analytical and
numerical results has been achieved. The lifetime of the skyrmion
scales with its initial size, $\lambda_0$, as
$(\lambda_{0}/a)^{5}$ in ferromagnets and as
$(\lambda_{0}/a)^{2.15}$ in antiferromagnets, with $a$ being the
lattice parameter. This makes antiferromagnetic skyrmions
significantly shorter lived than ferromagnetic skyrmions.
\end{abstract}


\maketitle Skyrmions \cite{SkyrmePRC58,BelPolJETP75} are
topologically stable configurations of a fixed-length
three-component vector field ${\bf n}({\bf r})$ in the coordinate
space of two dimensions. Due to the constraint ${\bf n}^{2}=1$ the
${\bf n}$-field has two independent components. This permits
unique mappings of ${\bf n} = (n_x,n_y,n_z)$ onto ${\bf r}=(x,y)$,
described by classes of homotopy \cite{Polyakov-book}. Each
homotopy class corresponds to a non-trivial field configuration
characterized by a conserved topological charge. The emergence of
a conserved charge from a continuous field theory prompted
numerous studies of skyrmions in problems of high-energy and
condensed matter physics \cite{Brown-book}.
They include cosmology \cite{DurKunMelPR02}, Bose-Enstein
condensates \cite{AlkStoNat01}, quantum Hall effect
\cite{SonKarKivPRB93,StonePRB93} and anomalous Hall effect
\cite{YeKimPRL99}, liquid crystals \cite{WriMerRMR89}.

The interest to
skyrmions in ordered spin systems had received much attention soon
after the discovery of high-temperature superconductivity in
copper oxides
\cite{WiegmannPRL88,ShrSigPRL88,WenZeePRL88,ChaHalNelPRB89,VorDonPRB90,GoodingPRL91,HaasPRL96},
and further explored recently
\cite{MarNet-PRB01,Mor-PRB02,Mor-PRB05,NazSan-PRL06}. It is
related to the fact that superconductivity in copper oxides occurs
in doped CuO$_{2}$ layers that, when undoped, are square lattices
of antiferromagnetically ordered spins. Initially there was some
hope that interaction of electrons and holes with skyrmions could
play some role in Cooper pairing but this was never successfully
demonstrated. Some indirect evidence of skyrmions in the
magnetoresistence of lanthanum copper oxide has been recently
reported \cite{Raicevic} but direct observation of skyrmions in
2$d$ antiferromagnetic lattices is still lacking.

In a continuous field model like, e.g., the non-linear
$\sigma$-model, the ground-state energy of the skyrmion does not
depend on its size, $\lambda$. This follows from the invariance of
the model with respect to the scale transformation ${\bf
r}\rightarrow k{\bf r}$, where $k$ is an arbitrary constant. If
the skyrmion  lives on a lattice, however, the scale
invariance becomes broken due to the presence of a lattice
parameter $a$. Thus the energy of the skyrmion depends on its size. This, in general,
must lead to the collapse or expansion of the skyrmion, making it
unstable.
The nature of the exchange interaction on a lattice makes the skyrmion energy decreasing with its size,
that leads to skyrmion collapse.
A number of authors looked for
interactions that could stabilize skyrmions in 2$d$ ferromagnets
\cite{AbanovPRB98,IvanovPRB06,IvanovPRB09}. It was argued that anisotropic
crystal field added to the isotropic exchange model may, in
principle, dynamically stabilize the skyrmion. In reality, however,
anisotropic interactions are of relativistic origin, while the
lattice effect that leads to the collapse of the
skyrmion is of the exchange origin and thus much greater. Therefore, it is important first
to understand what is the mechanism of skyrmions collapse in a generic exchange
model.

In this Letter we study the dynamics of skyrmions and the dependence of their collapse time $t_c$ on their
initial size in a $2d$ square lattice of classical spins coupled via
Heisenberg ferromagnetic (FM) or antiferromagnetic (AFM) exchange
interaction. The accuracy of the continuous approximation
increases with the size of the skyrmion, $\lambda$. One should,
therefore, expect that the lattice skyrmion becomes stable in the
limit of $\lambda\rightarrow\infty$. We find that $t_c$ of
the AFM skyrmion  scales as $t_c\propto(\lambda_{0}/a)^{2.15}$ with its initial size $\lambda_{0}$.
We compute the dynamics of the collapse
using both the analytical field model for the Neel vector and
a direct numerical calculation on lattices of up to 2000$\times$2000
exchange-coupled spins. The two approaches show
excellent agreement with each other. For a 2$d$ ferromagnet we
obtain (up to logarithmic corrections) the $(\lambda_{0}/a)^{5}$
scaling of the lifetime. This makes skyrmions significantly
shorter lived in a 2$d$ AFM than in a 2$d$ FM.

We begin with an antiferromagnet described by the Hamiltonian for
the Néel vector ${\bf L}$:
\begin{equation}
{\cal {H}}_0 =\frac{1}{2}JS^{2}\int
dxdy\left[\frac{1}{c^{2}}\dot{\bf L}^{2}+(\nabla{\bf
L})^{2}\right]\,.\label{H-L}
\end{equation}
Here ${\bf L}$ is normalized as ${\bf L}^{2}=1$, $(\nabla{\bf
L})^{2}\equiv(\partial_{x}{\bf L})^{2}+(\partial_{y}{\bf L})^{2}$,
$JS^{2}>0$ is the exchange energy associated with the interaction
of spins of length $S$, and $c$ is the speed of AFM spin waves
that equals $2\sqrt{2}Ja/\hbar$ in a square lattice.
The term with $\dot{\mathbf{L}}^{2}$ can be understood as a kinetic energy
responsible for the inertia of antiferromagnets.

The absolute
minimum of the energy corresponds to the uniform AFM background,
${\bf L}=\mathrm{const}$. Non-uniform configurations of ${\bf L}$
are characterized by the topological charge
\begin{equation}
Q=\frac{1}{4\pi a^{2}}\int dxdy\,{\bf L}\cdot(\partial_{x}{\bf
L}\times\partial_{y}{\bf L})\label{Q}
\end{equation}
that takes values $Q=0,\pm1,\pm2,\ldots$. Within, e.g., the
homotopy class $Q=-1$ the minimum energy, static configuration is
a skyrmion given by
\begin{equation}
{\bf L}=\left(\frac{2\lambda x}{r^2+\lambda^{2}},\,\frac{2\lambda
y}{r^2+\lambda^{2}},\,
\frac{r^2-\lambda^{2}}{r^2+\lambda^{2}}\right)\,,
\label{sn}
\end{equation}
where $r^2 = x^2 + y^2$. Its energy, $E=4\pi JS^{2}$, is
independent of $\lambda$.

\begin{figure}
\begin{centering}
\includegraphics[width=8cm]{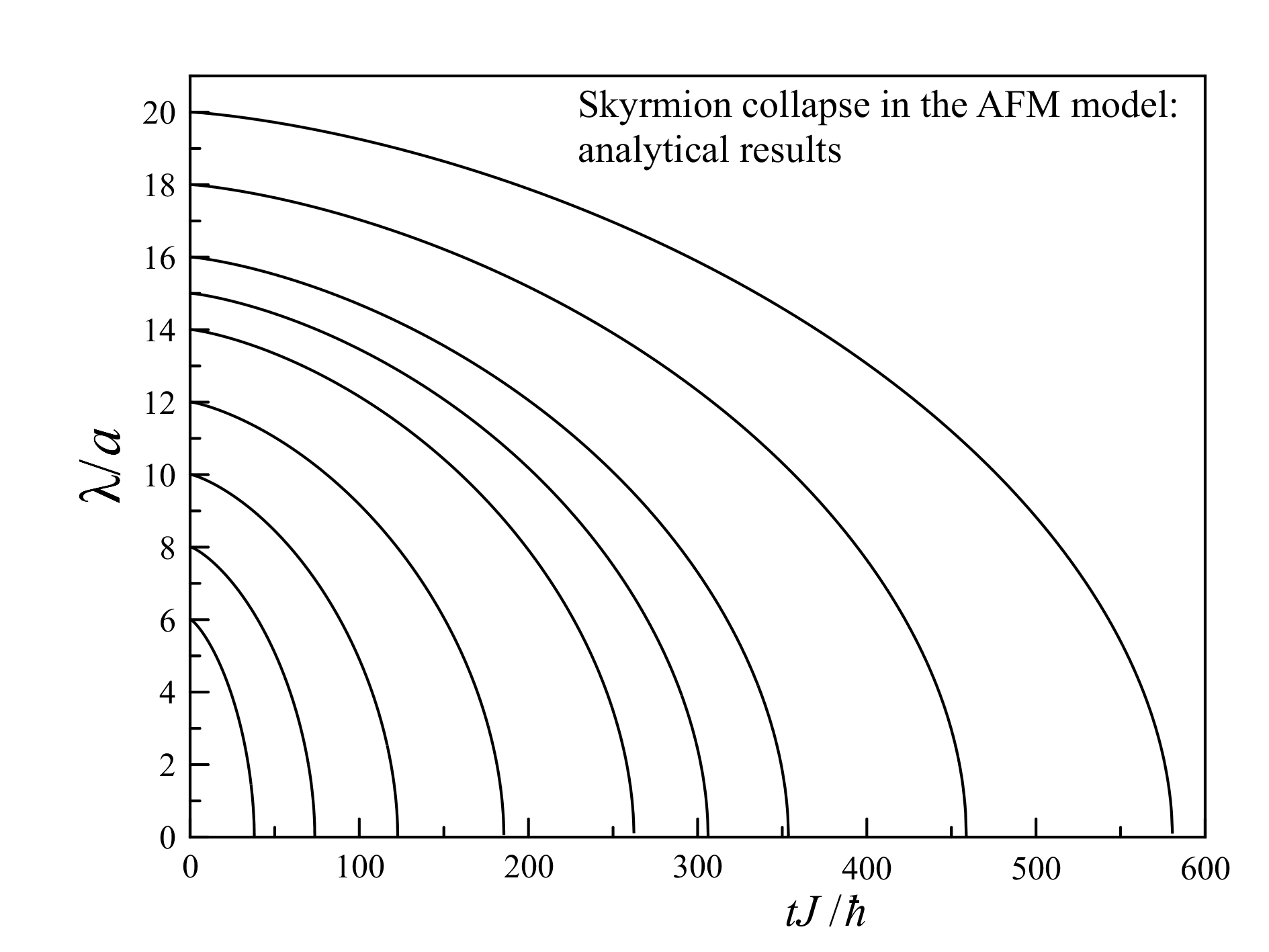}
\par\end{centering}
\caption{Collapse of antiferromagnetic skyrmions as described by the numerical
solution of Eq. (\ref{eq:lambda-Eq-AF}).\label{fig:Collapse-AF-analytical}}
\end{figure}

Equation (\ref{H-L}) can be derived from the Heisenberg exchange
interaction between nearest-neighbor classical spins
$\left|\mathbf{s}^{A}\right|=\left|\mathbf{s}^{B}\right|=1$,
\begin{equation}
\mathcal{H}=S^{2}\sum_{ij}J_{ij}\mathbf{s}_{i}^{A}\mathbf{\cdot s}_{j}^{B}
=-\frac{S}{2}\sum_{i\subset A}\mathbf{s}_{i}^{A}\mathbf{\cdot H}_{i}^{A}-
\frac{1}{2}\sum_{j\subset B}\mathbf{s}_{j}^{B}\mathbf{\cdot H}_{j}^{B}\,,\label{eq:Ham-AF}
\end{equation}
 where $A$ and $B$ denote AFM sublattices and $\mathbf{H}_{i}^{A,B}=-{\delta\mathcal{H}}/{\delta(S\mathbf{s}_{i}^{B,A})}=
 -S\sum_{ij}J_{ij}\mathbf{s}_{j}^{B,A}$ are the effective
 fields acting on the spins. As spins in each sublattice rotate smoothly through
space, one can expand the effective fields as
\begin{equation}
\mathbf{H}_{i}^{A}=-JS\left[4\mathbf{s}_{i}^{B}+a^{2}\nabla^{2}\mathbf{s}_{i}^{B}+\frac{a^{4}}{12}\left(\partial_{x}^{4}+\partial_{y}^{4}\right)\mathbf{s}_{i}^{B}+\ldots\right]\label{eq:HExpansion-AF}
\end{equation}
 and similar for $\mathbf{H}_{i}^{B}$. This allows one to go over to
the continuum description in which there are two spin fields
$\mathbf{s}^{A}$ and $\mathbf{s}^{B}$. Switching to the
magnetization
$\mathbf{M}=\left(\mathbf{s}^{A}+\mathbf{s}^{B}\right)/2$ and the
Néel vector
$\mathbf{L}=\left(\mathbf{s}^{A}-\mathbf{s}^{B}\right)/2$,
satisfying $\mathbf{M}^{2}+\mathbf{L}^{2}=1$ and $\mathbf{M\cdot
L}=0$,  with the help of  equations of motion $\hbar
\mathbf{\dot{s}}^{A,B}=\left[\mathbf{s}^{A,B}\times\mathbf{H}^{A,B}\right]$
 one obtains
\begin{equation}
\mathcal{H} = \mathcal{H}_0 -\frac{1}{24}JS^2a^{2}\int
dxdy\left[\left(\partial_{x}^{2}\mathbf{L}\right)^{2}+\left(\partial_{y}^{2}\mathbf{L}\right)^{2}\right]\,,
\label{eq:Ham-AF-Continuous-Ldot}
\end{equation}
which differs from Eq.\ (\ref{H-L}) by the second term due to the
discreteness of the lattice.
If the size of the skyrmion $\lambda$
is large compared to $a$, this term can be treated as a
perturbation.
Using the "rigid" skyrmion profile of Eq.\ (\ref{sn}), one obtains the energy due to this
term
\begin{equation}
{\cal E}_{\rm discr}= -(2\pi J S^2/3)(a/\lambda)^2
\label{EDiscr}
\end{equation}
that violates the scale invariance of the skyrmion.
Eq.\ (\ref{EDiscr}) can be interpreted as a potential energy
responsible for the skyrmion collapse.  During the collapse it is
transformed into the kinetic energy defined by the integral of
$\dot{\mathbf{L}}^{2} =
{4r^{2}}{\left(r^{2}+\lambda^{2}\right)^{-2}}\dot{\lambda}^{2}$.
 With account of energy conservation, Eq.\
(\ref{eq:Ham-AF-Continuous-Ldot}) gives
\begin{equation}
\frac{3}{c^{2}}\left(\ln\frac{r_{\max}^{2}+\lambda^{2}}{\lambda^{2}}-\frac{r_{\max}^{2}}{r_{\max}^{2}+\lambda^{2}}\right)\dot{\lambda}^{2}=\left(\frac{a}{\lambda}\right)^{2}-\left(\frac{a}{\lambda_{0}}\right)^{2},
\label{eq:lambda-Eq-AF}
\end{equation}
 where $\lambda_{0}$ is the initial size of the skyrmion and $r_{\max}$
has been introduced because of the logarithmic divergence of the integral in the kinetic energy.
The natural choice is $r_{\max}=\lambda_{0}+ct$, which
describes a front of AFM spin waves propagating away from the collapsing
skyrmion. This is confirmed by direct numerical calculations, see
Fig.\ \ref{fig:AF-front-propagation} below. The logarithmic terms
with time-dependent $r_{\max}$  require
numerical integration of Eq.\ (\ref{eq:lambda-Eq-AF}). The resulting collapse curves are shown in
Fig.\ \ref{fig:Collapse-AF-analytical}.

We now turn to the direct numerical solution of the dynamics of
the skyrmion provided by the microscopic Hamiltonian
(\ref{eq:Ham-AF}). The dynamics is determined by the coupled
equations of motion for spins, $\hbar
\mathbf{\dot{s}}_i=-\left[\mathbf{s}_i\times{\delta\mathcal{H}}/{\delta(S\mathbf{s}_{i})}\right]$.
We chose initial state as a staggered skyrmion texture, ${\bf
s}^{\rm st}$, given by Eq.\ (\ref{sn}) for the $A$ sublattice and
by the same formula but with a minus sign for the $B$ sublattice.
The size of the skyrmion numerically can be defined
as $\lambda^2_m =(m-1)(2^m\pi)^{-1}\sum_{i}\left(1-s_{zi}^{\rm
st}\right)^m$, where $m > 1 $ is an integer. If one replaces
summation by integration over $dxdy/a^2$ and uses the skyrmion
texture (\ref{sn}) for $s_z^{\rm st}$, this formula becomes an
identity, $\lambda_m = \lambda$.  The results presented below have been
obtained with $m=4$. Other options make little difference.

As the dynamics of the skyrmion is entirely due to small terms
arising from the lattice discreteness, the time dependence is slow
and sufficient accuracy can be achieved even for a large time step
of integration. Increasing the step is limited by stability rather
then by required accuracy. The challenge of the numerical solution
is the $1/r$ decay of the skyrmion profile that requires rather
big lattice sizes even for moderate values of $\lambda/a$. Free or
periodic boundary conditions introduce spurious
$\lambda$-dependent energies that compete with the small energy
due to the lattice discreteness, leading to the expansion of the
skyrmion instead of collapse. To make boundary conditions more
resembling an infinite lattice, we have included the missing
outside neighbors of the boundary spins with the values
approximated by the second-order extrapolation from the inside of
the working region. Still, the lattice size has to be large:
1000$\times1000$ for $\lambda_{0}/a$ up to 16 and $2000\times2000$
for $\lambda_{0}/a=18$ and 20. The program was implemented in
Wolfram Mathematica with a compiled vectorized fixed-step
fourth-order Runge-Kutta routine. One AFM skyrmion-collapse event required about one hour
computer time.

\begin{figure}
\begin{centering}
\includegraphics[width=8cm]{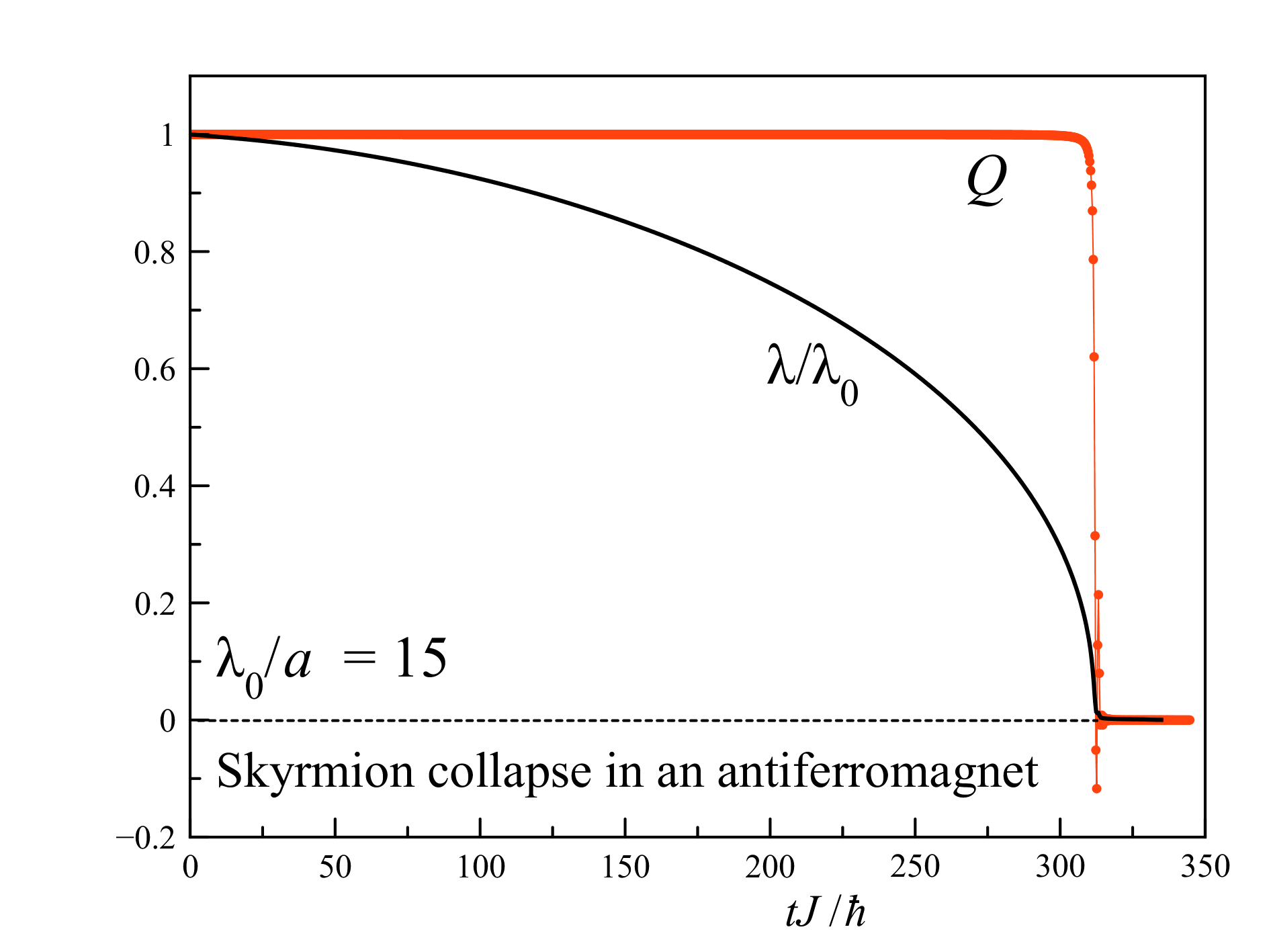}
\par\end{centering}
\caption{Skyrmion collapse in an antiferromagnet. Whereas the
skyrmion size $\lambda$ is decreasing continuously, the
topological charge $Q$ decays only during a short final stage of
the collapse.\label{fig:Skyrmion-collapse-AFM-Q}}
\end{figure}
\begin{figure}
\begin{centering}
\includegraphics[width=8cm]{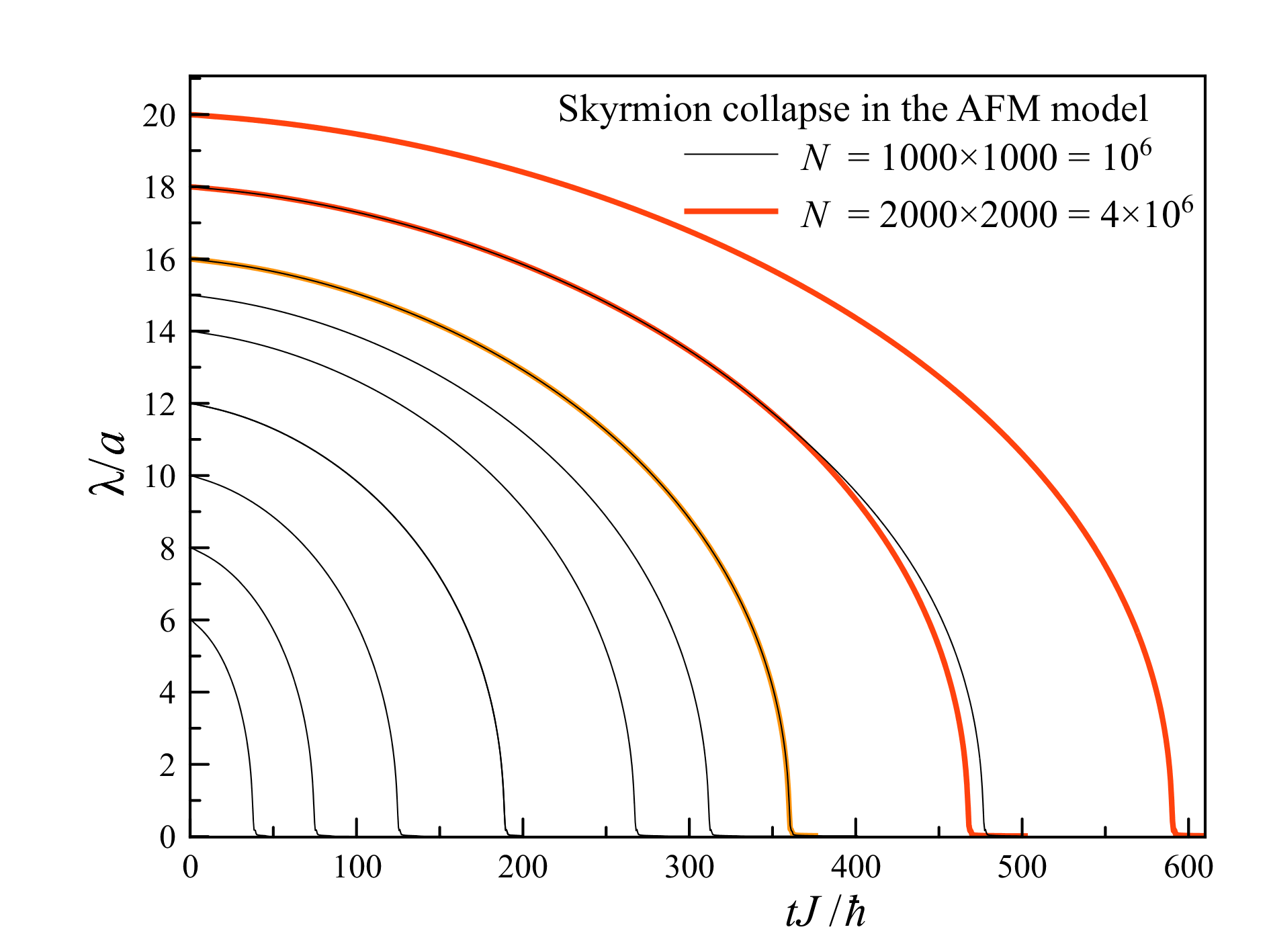}
\par\end{centering}
\caption{Skyrmion collapse in an antiferromagnet for different initial skyrmion
sizes.\label{fig:Skyrmion-collapse-AF}}
\end{figure}
The collapse of an AFM skyrmion with $\lambda_{0}/a=15$ is shown
in Fig. \ref{fig:Skyrmion-collapse-AFM-Q}. Whereas the skyrmion
size $\lambda$ is decreasing continuously, the topological
invariant $Q$ changes only during a short final stage of the
collapse when the continuous approximation fails. Fig.
\ref{fig:Skyrmion-collapse-AF} shows skyrmion collapse curves for
different values of $\lambda_{0}/a$. For $\lambda_{0}/a=18$ the
lattice size of one million spins is too small and computation
with four millions of spins is needed. For $\lambda_{0}/a=16$
these both lattice sizes yield the same collapse curve. These
results compare very well with the semi-analytical results shown
in Fig. \ref{fig:Collapse-AF-analytical}. The collapse time can be
fitted as $t_{c}\propto\lambda_{0}^{2.15}$ in this range of
$\lambda_{0}$. The considerable deviation from the square law can be traced back
 to the logarithmic term in Eq. (\ref{eq:lambda-Eq-AF}).
\begin{figure}
\begin{centering}
\includegraphics[width=8cm]{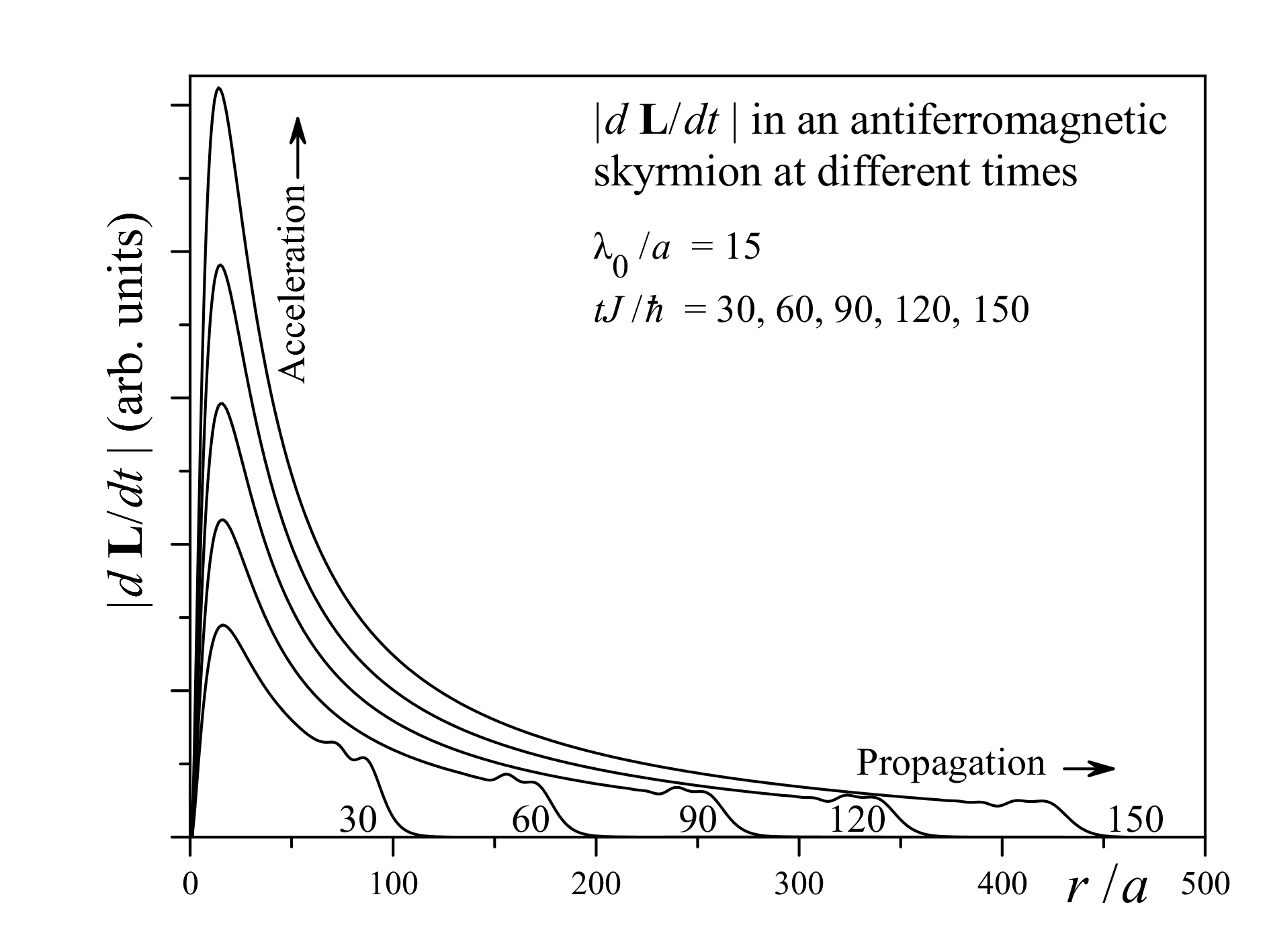}
\par\end{centering}
\caption{The front propagating from the center of
antiferromagnetic skyrmion at the beginning of its
collapse.\label{fig:AF-front-propagation}}
\end{figure}
Fig. \ref{fig:AF-front-propagation} shows
$\left|d\mathbf{L}/dt\right|$ in an antiferromagnetic skyrmion at
different times. The region of skyrmion motion where
$\left|d\mathbf{L}/dt\right|>0$ is expanding with the speed of
antiferromagnetic spin waves $c$. The reason for this is that the
lattice-discreteness terms that drive the skyrmion collapse are
very short ranged while the skyrmion itself is long-ranged. The
action of the former is transferred to the whole skyrmion with a
speed $c$ in accordance with the causality. The front position can
be estimated as $r_{\max}=\lambda_{0}+ct$, as was argued after Eq.
(\ref{eq:lambda-Eq-AF}).

\begin{figure}
\begin{centering}
\includegraphics[width=8cm]{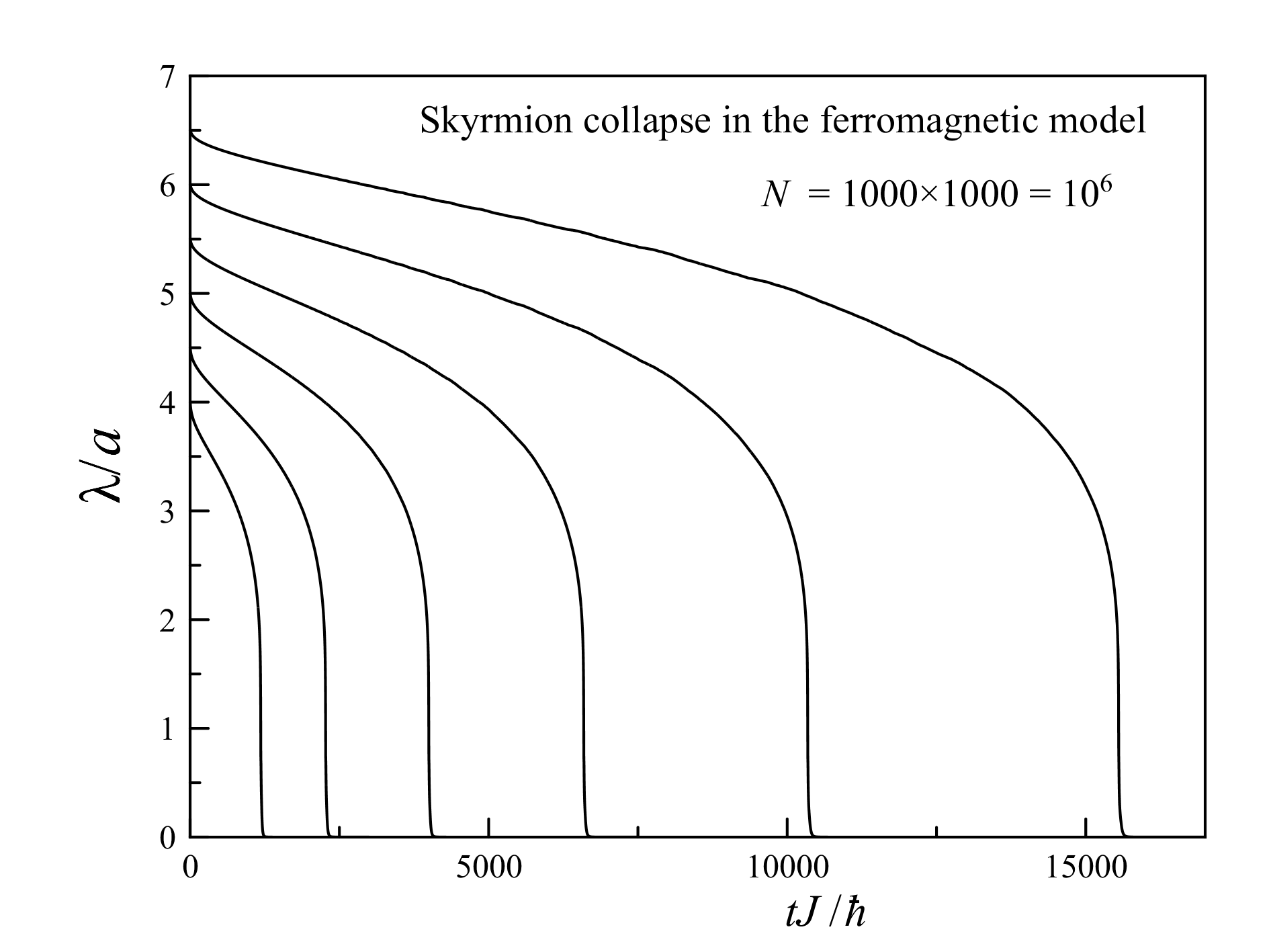}
\par\end{centering}
\caption{Skyrmion collapse in a ferromagnet. The collapse time scales as $t_{c}\propto\lambda_{0}^{5}$.\label{fig:F-skyrmion-collapse}}
\end{figure}
Along the same lines we have numerically studied the dynamics of ferromagnetic
skyrmions. It turns out to be much slower than the collapse of AFM skyrmions, so that up to one day
of computations is needed for one collapse event.
Fig. \ref{fig:F-skyrmion-collapse} shows time dependences of the
size of FM skyrmions during the collapse. The collapse time scales as $t_{c}\propto\lambda_{0}^{5}$.

%
The $\lambda_{0}^{5}$ scaling of the collapse time of the FM skyrmion can be
qualitatively understood as follows. The
exchange interaction conserves the total spin of the system. The infinitesimal increase of the
(negative) skyrmion spin in the course of its collapse is
\begin{equation}
d{\cal S} =  S\int\frac{d^2r}{a^2}\frac{ds_z}{d\lambda}d\lambda
 =  -8\pi S \frac{\lambda d
\lambda}{a^2}\ln\frac{R}{\lambda},\qquad d\lambda>0.
\end{equation}
 Here we used $s_z$ in the skyrmion form given by Eq.\
(\ref{sn}) and introduced the long range cut-off $R$.
Because of the conservation of the total spin,
the increase of the skyrmion spin by $d{\cal S}$ generates $d{\cal S}$ magnons.
Since in this process the spin is being carried by large distances,
the skyrmion collapse is very slow.
The average energy of emitted magnons can be estimated as
$\hbar \omega \sim -\hbar\dot \lambda/a$. This yields the
emitted magnon power
\begin{equation}\label{P1}
P = \hbar\omega \frac{d{\cal S}}{dt} = 8\pi \hbar
S\frac{\lambda\dot \lambda^2}{a^3}
\ln\frac{R}{\lambda}\,.
\end{equation}
On the other hand, the rate of change of the energy (\ref{EDiscr})
due to discreteness of the lattice is $\dot{\cal E}_{\rm discr}
\propto \dot\lambda$. From the energy conservation, $\dot{\cal
E}_{\rm discr}+P=0$, one obtains
\begin{equation}\label{derivative}
\frac{d\lambda}{dt} = -\frac{JSa^5}{6\hbar}\cdot
\frac{1}{\lambda^4\ln(R/\lambda)}\,,
\end{equation}
yielding the collapse time
\begin{equation}\label{tc-FM}
{t}_c = \frac{6\hbar}{5JS}\left(
\frac{{\lambda}_0}{a}\right)^5\ln\left(\frac{{R}}{{\lambda}_0}\right)\,.
\end{equation}
The condition $\hbar\omega \ll SJ$ for the energy of the magnons
translates to $5({\lambda}/{a})^4\ln({R}/{\lambda}) \gg 1$, which
is well satisfied during the collapse.

In conclusion, we have studied the collapse of skyrmions due to
the discreteness of the lattice in generic models of isotropic
2$d$ ferro- and antiferromagnets with Heisenberg exchange
interaction. The results obtained within continuous field model
are in excellent agreement with the direct numerical calculation
on lattices of up to 2000$\times$2000 coupled spins. The collapse
time of antiferromagnetic skyrmions obtained by both methods
scales as $(\lambda_{0}/a)^{2.15}$. For ferromagnetic skyrmions,
the numerical calculation gives the $(\lambda_{0}/a)^{5}$ scaling
of the collapse time. It is explained by the emission of magnons.
Thus, AFM skyrmions are much shorter lived than FM skyrmions. This
can be understood in the following terms. The collapse of an AFM
skyrmion occurs via transformation of its potential energy due to
the discreteness of the lattice into the kinetic energy defined by
$\dot{\bf L}^2$. The FM skyrmion does not possess such a kinetic
energy, so that its potential energy has to be dissipated into
magnons, which is a much slower process. In the expression for
$t_c$ the time constant in front of the power of the ratio
$\lambda_0/a$ is of order $\hbar/(JS)$. For, e.g., $JS \sim 100K$
and $\lambda_0 \sim 10a$, this gives $t_c \sim 10$ns for the
lifetime of the skyrmion in a ferromagnet and $t_c \sim 10$ps in
an antiferromagnet.

The authors thank Oliver R\"ubenk\"onig and Daniel Lichtblau of
Wolfram Research for helping with vectorization and compilation
in Wolfram Mathematica.
This work has been supported by the Department of Energy through grant
No. DE-FG02-93ER45487.


\begin{thebibliography}{10}
\bibitem{SkyrmePRC58} T. H. R. Skyrme, Proc. Roy. Soc. London, Ser.
A \textbf{247}, 260 (1958).

\bibitem{BelPolJETP75} A. A. Belavin and A. M. Polyakov, Pis'ma Zh.
Eksp. Teor. Fiz \textbf{22}, 503 (1975) {[}JETP Lett. \textbf{22},
245 (1975).

\bibitem{Polyakov-book} A. M. Polyakov, \textit{Gauge Fields and
Strings}, Harwood Academic Publishers 1987.

\bibitem{Brown-book} {\it The Multifaceted Skyrmion}, edited
by G. E. Brown and M. Rho (World Scientific, 2010).

\bibitem{DurKunMelPR02} R. Durrer, M. Kunz, and A. Melchiorri, Phys.
Rep. \textbf{364}, 1 (2002).

\bibitem{AlkStoNat01} U. Al'Khawaja, and H. T. C. Stoof, Nature \textbf{411},
918(2001).

\bibitem{SonKarKivPRB93} S. L. Sondhi, A. Karlhede, S. A. Kivelson,
and E. H. Rezayi, Phys. Rev. B \textbf{47}, 16419 (1993).

\bibitem{YeKimPRL99} Jinwu Ye, Y. B. Kim, A. J. Millis, B. I. Shraiman,
P. Majumdar, and Z. Tesanovic, Phys. Rev. Lett. \textbf{83}, 3737
(1999).

\bibitem{StonePRB93} M. Stone, Phys. Rev. B \textbf{53}, 16573 (1996).

\bibitem{WriMerRMR89} D. C. Wright, and N. D. Mermin, Rev. Mod. Phys.
\textbf{61}, 385 (1989).

\bibitem{WiegmannPRL88} P. B. Wiegmann, Phys. Rev. Lett. \textbf{60},
821 (1988).

\bibitem{ShrSigPRL88} B. I. Shraiman and E.D. Siggia, Phys. Rev.
Lett. \textbf{61}, 467 (1988).

\bibitem{WenZeePRL88} X. G. Wen and A. Zee, Phys. Rev. Lett. \textbf{61},
1025 (1988).

\bibitem{HaldanePRL88} F. D. M. Haldane, Phys. Rev. Lett. \textbf{61},
1029 (1988).

\bibitem{ChaHalNelPRB89} S. Chakravarty, B. I. Halperin, and D. R.
Nelson, Phys. Rev. B \textbf{39}, 2344 (1989).

\bibitem{VorDonPRB90} P. Voruganti and S. Doniach, Phys. Rev. B \textbf{41},
9358 (1990).

\bibitem{GoodingPRL91} R. J. Gooding, Phys. Rev. Lett. \textbf{66},
2266 (1991).

\bibitem{HaasPRL96} S. Haas, F.-C. Zhang, F. Mila, and T. M. Rice,
Phys. Rev. Lett. \textbf{77}, 3021 (1996).

\bibitem{MarNet-PRB01} E. C. Marino and M. B. Silva Neto, Phys. Rev.
B \textbf{64}, 092511 (2001).

\bibitem{Mor-PRB02} T. Morinari, Phys. Rev. B \textbf{65}, 064513
(2002).

\bibitem{Mor-PRB05} T. Morinari, Phys. Rev. B \textbf{72}, 104502
(2005).

\bibitem{NazSan-PRL06} Z. Nazario and D. I. Santiago, Phys. Rev.
Lett. \textbf{97}, 197201 (2006).

\bibitem{Raicevic} I. Raicevic, D. Popovic, C. Panagopoulos, L. Benfatto,
M. B. Silva Neto, E. S. Choi, and T. Sasagawa, Phys. Rev. Lett. \textbf{106},
227206 (2011).

\bibitem{AbanovPRB98} A. Abanov and V. L. Pokrovsky, Phys. Rev. B
\textbf{58}, R8889 (1998).

\bibitem{IvanovPRB06} B. A. Ivanov, A. Y. Merkulov, V. A. Stepanovich,
C. E. Zaspel, Phys. Rev. B \textbf{74}, 224422 (2006).

\bibitem{IvanovPRB09} E. G. Galkina, E. V. Kirichenko, B. A. Ivanov,
V. A. Stephanovich, Phys. Rev. B \textbf{79}, 134439 (2009).\end{thebibliography}
\end{document}